\documentclass[twocolumn,prb,amsmath,amssymb]{revtex4}
\usepackage{graphicx}
\usepackage{bm}
\begin{document}

\title{Sensitive detection of photoexcited carriers by resonant tunneling\\ through a single quantum dot}

\author{E.~E.~Vdovin$^{1,2}$, O.~N.~Makarovsky$^{1}$, A.~Patan\`e$^{1}$\footnote{Electronic e-mail: amalia.patane@nottingham.ac.uk}, L.~Eaves$^{1}$, and Yu.~N.~Khanin$^{2}$}

\affiliation{$^{1}$~School of Physics and Astronomy, The University of Nottingham, NG7 2RD, UK
\\$^{2}$~Institute of Microelectronics Technology RAS, 142432 Chernogolovka, Russia}

\begin{abstract} 
We show that the resonant tunnel current through a single energy level of an individual quantum dot within an ensemble of dots is strongly sensitive to photoexcited holes that become bound in the close vicinity of the dot. The presence of these holes lowers the electrostatic energy of the quantum dot state and switches the current carrying channel from fully open to fully closed with a high on/off ratio ($> 50$). The device can be reset by means of a bias voltage pulse. These properties are of interest for charge sensitive photon counting devices. 
\end{abstract}

\maketitle

\section{INTRODUCTION}
Semiconducting quantum dots (QDs), formed by lithographic processing or by self-assembly, have proved to be versatile nanostructures for studying and exploiting resonant electron tunneling through a single, discrete quantum state \cite{1,2,3,4,5,6}. Usually, the resonance condition is achieved by tuning the voltage applied either to a two-terminal diode \cite{4,5,6,7} or else to a gate electrode of a transistor structure \cite{1,2,3,7}. The electrostatic potential arising from charges on a surface gate tends to have only a gradual spatial variation, given by Poisson's equation and characterized by a length scale that is large compared to the typical size ($\sim 10$ nm) of a QD. However, the presence of a single quantum of charge localized within a distance of 10 nm from a dot can change its Coulomb potential energy by several meV. Here we exploit this concept and show that the tunnel current through a single QD is strongly sensitive to a small number of photoexcited holes that become bound in the vicinity of the dot. In previous studies, the trapping of photoexcited charges on QDs has been used as a detector of single-photons or as a ``floating gate'' for counting the number of photons in a single pulse of light. However, in these experiments the step-like changes in the tunnel current through the quantum well of a resonant tunneling diode \cite{8} or in the conductivity of a two-dimensional (2D) electron gas \cite{9,10} were small, typically 1\% of the current, or less. Persistent photoconductivity effects caused by hole trapping effects have also been observed in studies of narrow one-dimensional (1D) constrictions in a 2D electron gas \cite{12}. In our work, we use devices in which the current over a narrow range of bias arises from resonant tunnelling through a single ``active'' QD state. The positive charge of the bound photoexcited holes shifts the resonance condition to lower applied bias thus providing a means of switching the current-carrying channel from fully open to fully closed, an effect which has potential for charge-sensitive photon counting detectors \cite{8,9,10,11}.

\begin{figure}[b]
\includegraphics[width=3.0 in]{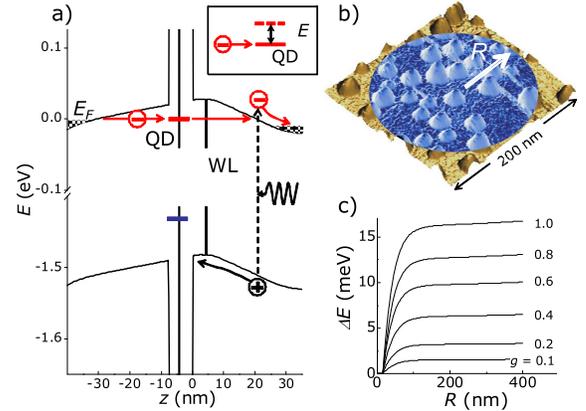}
\caption{(Colour online) a) Electrostatic potential energy profile for our device structure and sketch of an electron tunnelling from the n-GaAs emitter layer into the energy level of a single InAs QD. The inset shows the energy shift $\Delta E$ of the QD energy level due to the presence of photoexcited holes. b) Scanning tunnelling microscopy image of InAs QDs grown under similar conditions to our tunnel diodes. The circle highlights an area of radius $R \sim 100$ nm. c) Calculated energy shift  $\Delta E$ versus the radius $R$ of the QD ensemble for different values of g (g=1 corresponds to all dots filled with a single hole).}
\end{figure}

\section{DEVICES AND EXPERIMENTAL DATA}

\vspace{5.5mm}
In our tunnel diodes, a layer of self-assembled InAs quantum dots is incorporated in an Al$_{0.4}$Ga$_{0.6}$As tunnel barrier and gives rise to a planar ensemble of discrete atom-like states with energy levels close the conduction band minimum of GaAs \cite{13}. Under an applied bias electrons can tunnel into these QD states from the Fermi sea of an adjacent n-doped GaAs layer. A schematic band diagram of our device is shown in Figure 1a (see ref.14 for a detailed description of the device composition). A two-dimensional InAs wetting layer (WL) on the collector side of the barrier produces a depletion layer in close proximity to the InAs QDs, thus facilitating electron tunneling from the dots through the continuum states of the WL and finally into the n-doped GaAs collector contact, see Figure~1a. Despite the large number of QDs in our diode ($\sim 10^6$ for a 25~$\mu$m diameter mesa), only a small number of narrow peaks can be typically observed in the low bias ($<0.1$V) current-voltage curves, $I(V)$ \cite{4,5,6,7}. This behaviour is associated with a limited number of tunneling channels that can efficiently transmit electrons in this low bias range. These channels are influenced by the local and random distribution of Si-donor impurities in the nearby doped GaAs layers, by residual strain and by charging of the dots \cite{7}.

Figure 2a shows $I(V)$ curves measured at $T = 4.2$ K in the dark and under illumination using laser light of wavelength  $\lambda = 660$ nm. In the absence of light, a sharp peak is observed in $I(V)$ at $V \sim 20$ mV. Its form is approximately triangular with a sharp and temperature-dependent onset at low bias, which is consistent with energy-conserving resonant tunneling of electrons from a thermalised degenerate 3D Fermi gas in the GaAs emitter into a 0D QD state in the barrier \cite{14}. Thus, at this low bias, the tunnel current is determined by a single ``active'' QD. With laser illumination, the peak in $I(V)$ shifts to lower bias. This effect is accompanied by temporal fluctuations in the tunnel current, which exhibits step-like transitions between two or more discrete values, an effect generally referred to as telegraph noise \cite{15} (Figure~2b).

\begin{figure}[t]
\includegraphics[width=2.9 in]{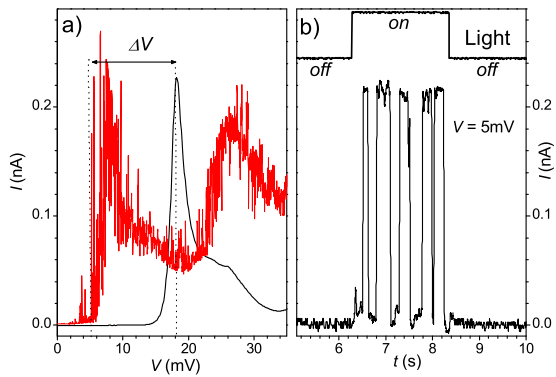}
\caption{(Colour online) a) $I(V)$ curve at $T = 4.2$ K in the dark (black line) and under illumination (red line) with laser light ( $\lambda = 660$ nm and $P= 0.1$ W/m$^2$). b) Time dependence of the tunnel current at $V= 5$ mV in the dark and under illumination. 
}
\end{figure}

With increasing temperature, the onset of the tunnel current though the active QD state broadens due to the thermal smearing ($\sim k_BT$) of the Fermi level in the emitter. Correspondingly, the light-induced bias shift,  $\Delta V$, of the peak in $I(V)$ and the telegraph noise are quenched with increasing $T$ and disappear at $T \sim 50$K. An increasing level of illumination causes a faster on/off switching of the tunnel current. Figure~3a shows the effect of increasing the illumination intensity on the switching of the current between its ``on'' and ``off'' states at an applied voltage of $V = 5$ mV, well below the onset of the resonant peak in the dark $I(V)$. Note that at higher illumination intensities the QD channel spends more time in the ``on'' mode. This is consistent with data in Figure~2, showing that the resonant peak in the time-averaged $I(V)$ curve shifts to lower biases under illumination. The bias shift, $\Delta V$,  of the peak in $I(V)$ is shown in Figure~3b as a function of the power, $P$, of the laser light ( $\lambda = 660$ nm). These effects were observed for photoexcitation of the diode over a wide range of wavelengths smaller than  $\lambda = 860$ nm ($hv = 1.44$ eV), the threshold condition for photon excitation of electron-hole pairs in the InAs WL between the tunnel barrier and the electron collector layer \cite{16}. Data similar to those in Figures~2-3 were observed in other tunnel diodes incorporating InAs quantum dots, which also revealed sharp QD resonances just above the threshold voltage for current flow.

\begin{figure}[t]
\includegraphics[width=3.6 in]{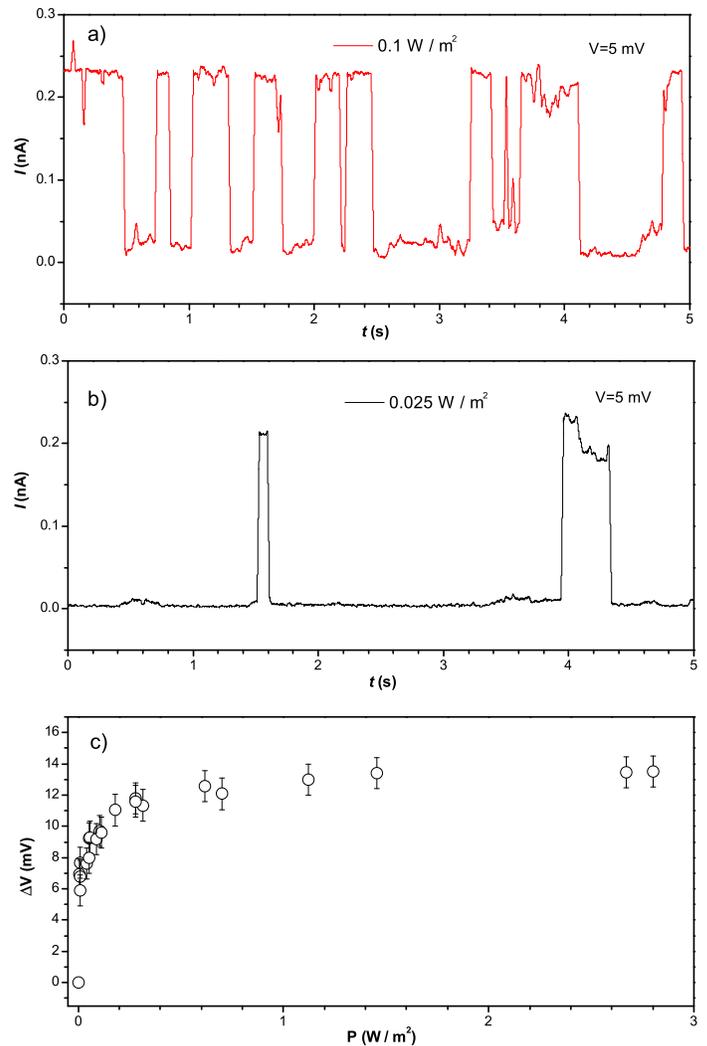}
\caption{(Colour online) a-b) Time dependence of the current under illumination with laser light ( $\lambda = 660$ nm, $V= 5$ mV and $T = 4.2$ K). c) $P$-dependence of the light-induced voltage shift,  $\Delta V (\sim \Delta E)$ of the peak in $I(V)$. 
}
\end{figure}

The response of the tunnel current to illumination is further illustrated in Figure 4a. Here we set the applied bias at $V = 5$~mV, well below the onset of the resonant peak in $I(V)$ in the dark. Following excitation of the diode with a short ($< 10$ ms) light pulse from a laser diode ( $\lambda = 660$ nm), the tunnel current increases to its resonant peak value of 0.2~nA and then falls to zero, through either single- or multiple-steps over long ($> 0.1$s) time intervals. To monitor the corresponding temporal evolution of the QD resonance, we measure the $I(V)$ curve in a fast acquisition mode ($\sim 10$ ms) and at regular intervals of time. Following illumination, measurements of the resonant peak in $I(V)$ in the fast mode are free of telegraph noise and, with time, the peak shifts from ~ 5~mV to higher biases, back towards its voltage position in the dark, see Figure~4b. The $I(V)$ curve can be restored to its initial ``dark'' state by applying a short negative bias reset pulse. This causes a short discharging current pulse after which the system is ready to detect the next optical pulse, see Figure~4c.

\section{DISCUSSION}
To explain these data, we consider the effect of illumination on an electron tunneling through a single QD. As sketched in Figure~1a, the light creates photocarriers (electrons and holes) in the WL and the undoped GaAs layers on either sides of the tunnel barrier. The photoelectrons are swept by the electric field into the electron collector and have little effect on the device properties, whilst the holes move towards the QD layer, where they are captured by the dots. The positive charge surrounding the active QD shifts the energy of its resonant tunneling level downwards by an amount  $\Delta E$. Using the electrostatic leverage factor, $f = 0.44 \pm 0.05$, which gives the fraction of the applied voltage dropped between the Fermi energy of the emitter $E_F$ and the QD state, we determine  $\Delta E$ from the photo-induced voltage shift, $\Delta V$, i.e.  $\Delta E = fe \Delta V$ \cite{17}. The measured shift,  $\Delta E$, of up to 6 meV, is comparable with the Coulomb interaction energy of an electron bound in a QD with a single hole at a distance of 20~nm. This is close to the average spatial separation (30~nm) between the dots in our devices (Figure~1b), thus indicating that the photoexcited holes, which influence the active QD, are localised at adjacent dots of the QD layer; alternatively, they are captured in the potential minima associated with dot-related residual strain in the nearby InAs WL. 

\begin{figure}[t]
\includegraphics[width=3.0 in]{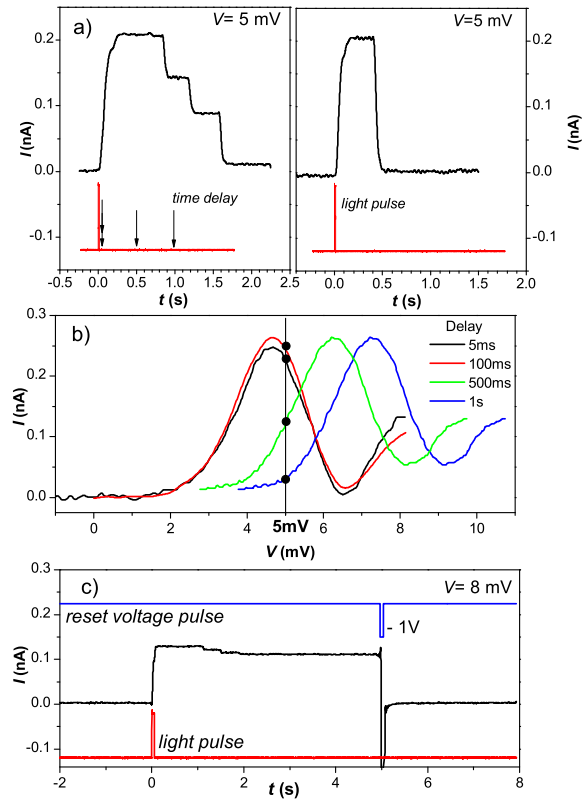}
\caption{(Colour online) a) Time dependences of the tunnel current at $V = 5$ mV following excitation of the diode with a short (10~ms) light pulse. b) $I(V)$ curves acquired in the fast mode and at given delay times following excitation of the diode with a short (10~ms) light pulse. c) Effect of a short light pulse and of a negative bias ``reset'' pulse on the time dependence of the tunnel current at $V = 8$ mV. In parts (a-c) $T = 4.2$K and the diode is excited with light from a laser diode with  $\lambda = 660$~nm and $P = 3$~W/m$^2$. 
}
\end{figure}

Under constant illumination, the charging and discharging of the dots by the photoexcited holes lead to temporal fluctuations in the bias position of the QD resonance. This results in an on/off switching of the tunnel current measured at bias values $V < V_d$, where $V_d$ is the bias condition for the QD resonant peak in the dark $I(V)$ (Figures~2 and 3a). For an increasing intensity of illumination and at $V < V_d$, the active QD spends more time in the ``on'' mode due to the higher occupancy of the QD states with holes (Figure~3a). To examine this phenomenon in the time-domain, we consider the temporal variation of the tunnel current (Figure~4a) and of the resonant peak in $I(V)$ (Figure~4b) following a short light pulse excitation. The time-dependences shown in Figure~4a and 4b are determined by the discharging of the dots. Due to the decay of the Coulomb interaction at large distances ($> 100$~nm), it is reasonable to assume that only charged dots in close proximity to the active QD have a significant effect on the current channel. Since the number of nearby dots is small, their discharging leads to discrete step-like, rather than continuous, changes in the energy of the active QD level relative to the Fermi energy in the emitter. Hence, the bias condition for the QD resonance (Figure~4b) and the measured current at a particular fixed bias (Figure~4a) also change in discrete steps.
 
The temporal variation of the tunnel current depends on the intensity of the light pulse: multiple step-decreases of the current to zero are more frequently observed at higher levels of light intensity, whereas at low intensities a single step to zero current is most commonly observed. This indicates that at low illumination levels, the single step decrease to zero involves the discharge of a single QD close to the active QD. The detrapping of holes localized onto the quantum dots is a random quantum process. Figure 4c shows an event in which a photoexcited hole remains localized to the active QD for several seconds. The trace in this Figure also shows how we can use a short negative bias reset pulse to discharge this hole and restore the current from its ``on'' value back to zero.

The characteristic time for discharge of the dots,  ${\tau} _d \sim 0.1$ s, is much longer than the characteristic dwell time ( ${\tau}_t = e/I < 1$ ns) of electrons in the active dot and the radiative time (${\tau}_ r < 1$ns) for electron-hole recombination in QDs. Therefore, direct exciton recombination in QDs plays no role in the slow discharging; such recombination takes place via non-radiative recombination or by spatially-indirect recombination of the photoexcited holes with electrons in the nearby GaAs layers \cite{18}. As can be seen in Figure 3b, for low levels of light illumination, the increase in  $\Delta V$ with $P$ indicates an increasing amount of hole charge trapped in the QDs. The voltage shift saturates at high powers. This is partially due to the dynamical balance between two competing processes: the filling of the dots by the photoexcited holes and the discharging of the dots by the incoming electrons from the emitter layer when the hole charging pushes the QD levels below the emitter Fermi energy. 

To quantify the discussion, we model the energy shift $\Delta E$ that results from a random and uniform distribution of hole charges bound by the dots surrounding an empty dot. The positions of the dots in their growth plane $xy$ and the filling of the dots with holes are simulated by sets of random numbers, giving an average occupancy of the dots, g ($0 \le$ g $\le 1$), which increases with increasing light intensity. The simulation uses an area of radius $R$ that varies from 15~nm  to 1 $\mu$m, corresponding to a number of dots ranging from 1 to $\sim 10^3$, respectively (dot density $n_{QD} = 10^{11}$ cm$^{-2}$), Figure~1b. It takes into account image charges generated by the photoexcited holes in the n-type GaAs contact layers on either sides of the QD plane and considers over 500 positions of the empty dot in the $xy$ plane. As shown in Figure~1c, $\Delta E$ increases with $R$ and saturates at large $R$ ($>100$ nm). This dependence reflects the rapid decay of the Coulomb interaction at large distances due to the screening effect of the nearby conducting n-type GaAs contact layers. For a given $R, \Delta E$ increases with g and reaches a maximum when all dots are filled with holes (g=1). The measured saturation value  $\Delta E_s = (6 \pm 1)$ meV correspond to the calculated values for g = 0.4. A simple calculation \cite{19} shows that this value of g is consistent with the slow discharging of the dots ( ${\tau}_d \sim 0.1$ s) and the number of photogenerated holes when  $\Delta E$ saturates to its largest value.
 
The capture of photoexcited carriers on QDs has been used recently as a ``floating gate'' in a photon-number discriminating detector \cite{10}: in that device the capture of a photo-excited carrier gives rise to very small step-like changes ($< 0.003$\%) in the conductivity of a two-dimensional electron gas. Single photon detection has also been achieved by using the trapping of photo-excited charges on QDs to induce small ($\sim 1$\%) steps in the resonant tunnelling current through the quantum well of a resonant tunnelling diode \cite{8}. In these previous studies, the trapped charge has only a local effect on the current flow, affecting a small area of the conducting region and changing the overall conductivity by only $\sim 1$\% or even less. In our devices, we can achieve a much higher level of sensitivity ($>1000$\%) since, over a limited range of bias, the current is due entirely to resonant tunnelling through a discrete energy level of a single active QD; the presence/absence of photoexcited holes localized in bound states within $\sim 100$nm of the active dot provides a means of opening/closing the current channel with a very high on/off ratio ($> 50/1$).

\section{CONCLUSION}
In conclusion, we have shown how the resonant tunnel current through a quantum energy level of an individual quantum dot within an ensemble of dots in a tunnel diode is highly sensitive to photoexcited holes that become bound in the close vicinity of the dot. The presence of these holes lowers the electrostatic energy of the active quantum dot tunnelling channel and, for a fixed applied bias, switches the channel from fully open to fully closed. This high sensitivity, combined with the ability to remove the holes with a voltage reset pulse, opens up prospects for exploitation of resonant tunneling through a quantum dot for charge-sensitive photon counting detectors.

\begin{acknowledgments}
The work is supported by EPSRC, the Royal Society (UK) and the RFBR (Russia). We acknowledge P.J.~Moriarty for providing the STM images of our quantum dots, M.~Henini for growing the layers and R.~Airey for processing the devices.
\end{acknowledgments}


\begin{thebibliography}{text}
\bibitem{17} The temperature dependence of the current onset allows us to determine the electrostatic leverage factor, $f$, which relates the applied voltage, $V$, to the energy of the dot state, $E$, measured relative to the Fermi energy in the emitter, i.e. $E =eVf$. Below threshold, the $T$- and $V$-~dependence of the tunnel current into a dot state of energy E  can be approximated by the Boltzmann tail of the Fermi-Dirac distribution, i.e. $I \propto \exp(-E/k_{B}T)=   \exp(-eVf/k_{B}T)$. The plot of $\ln(I)$ versus $eV$ is described by a straight line with a gradient given by $f/k_{B}T$. The leverage factor is derived by the slope of the measured gradient versus $(k_{B}T)^{-1}$.
\bibitem{19} We express the rate of photoexcited holes as $R_h=$g$n_{QD}/{\tau}_d$, where g is the average occupancy of the dots with holes. Here $n_{QD}=10^{11}$~cm$^{-2}$ is the density of QDs and  ${\tau}_d \sim 0.1$~s is the discharging time. The rate can be also expressed as $R_h=Pe^{-\alpha L}\alpha t/hv$, where $P$ is the power density, $hv =1.9$~eV is the photon energy, $\alpha = 2 \times 10^6$ m$^{-1}$ is the absorption coefficient of GaAs at $hv =1.9$~eV, $t  = 24$~nm is the thickness of the GaAs layer where electrons and holes are swept by the electric field and $L = 0.6 \mu$m is the distance of the QD layer from the surface. By equalizing the two expressions for $R_h$ , we derive g $= {\tau}_dPe^{-\alpha L}\alpha t/n_{QD}hv$ . At the highest level of illumination ($P > 0.1$ W/m$^2$), we find g $> 0.5$.
\end{thebibliography}
\end{document}